\documentclass[aps,prd,preprint,tightenlines,superscriptaddress]{revtex4}
\usepackage{epsfig}
\usepackage{graphicx}
\usepackage{psfrag}
\usepackage{amsmath,amssymb}
\usepackage{colordvi}
\usepackage{color}
\usepackage{amsfonts}
\usepackage{enumerate}
\usepackage{slashed}

\newcommand{\la}{\langle}
\newcommand{\ra}{\rangle}

\begin{document}

\title{Gluon Spin, Canonical Momentum, and Gauge Symmetry}
\author{Xiangdong Ji}
\affiliation{INPAC, Department of Physics, and Shanghai Key Lab for Particle Physics and Cosmology,
Shanghai Jiao Tong University, Shanghai, 200240, P. R. China}
\affiliation{Center for High-Energy Physics, Peking University, Beijing, 100080, P. R. China}
\affiliation{Maryland Center for Fundamental Physics, University of Maryland, College Park, Maryland 20742, USA}
\author{Yang Xu}
\affiliation{Center for High-Energy Physics, Peking University, Beijing, 100080, P. R. China}
\affiliation{Maryland Center for Fundamental Physics, University of Maryland, College Park, Maryland 20742, USA}
\author{Yong Zhao}
\affiliation{Maryland Center for Fundamental Physics, University of Maryland, College Park, Maryland 20742, USA}

\date{\today}
\vspace{0.5in}
\begin{abstract}
It is well known that in gauge theories, the spin (and orbital angular momentum)
of gauge particles is not gauge invariant, although the helicity is;
neither are the canonical momentum and canonical angular momentum of
charged particles. However, the simple
appeal of these concepts has motivated repeated attempts to resurrect them as physical
descriptions of gauge systems. In particular, measurability of the
gluon-spin-contribution to the proton helicity in polarized proton
scattering has generated many theoretical efforts in generalizing it
and others as gauge-invariant quantities. In this work, we analyze the constraints
of gauge symmetry, the significance of gluon spin in the light-cone gauge,
and what is possible and natural in QCD parton physics, emphasizing
experimental observability and physical interpretation in the structure of
bound states. We also comment on the measurability of the orbital angular momentum
of the Laguerre-Gaussian laser modes in optics.

\end{abstract}

\maketitle

\section{Introduction}

Feynman's parton is a powerful idea describing
high-energy collisions of hadrons~\cite{Feynman:1969ej}. The parton concept
grows out of a simple model for hadrons in the infinite momentum frame
(IMF), i.e., observers or systems traveling at the speed of light, in which the interactions
are slowed down by Lorentz dilation and the hadrons appear to be simply a collection
of non-interacting partons. When quantum chromodynamics (QCD) was proposed as the
fundamental quantum theory to describe the physics of strong interactions,
the parton picture survived with two minor modifications: First, it
emerges in a special choice of gauge, the light-cone gauge $A^+=0$ (where $+$ means a combination
$(A^0+A^3)/\sqrt{2}$ of $A^\mu$, $\mu=0,1,2,3$, if the hadron is moving in the 3-direction),
because the quarks do interact with each other through an SU(3) color
gauge field, even in the IMF. In this choice of gauge, the interaction effects
are nominally gone. Second, partons act differently at different (momentum) resolution scales,
again due to color interactions. However, the scale evolution
can be computed in QCD perturbation theory and is believed to be
under control~\cite{Sterman:1994ce}. Of course, QCD also permits studying effects beyond
the simple parton model by including parton correlations, transverse momentum, off-shellnesss, etc.
These effects cannot be studied model-independently, and one must
invoke high-twist factorization and expansions~\cite{Ellis:1982cd, Ellis:1982wd}.

Because the parton concept is so appealing, it often motivates
attempts to do away with the gauge symmetry of QCD and
search for ``physical observables", which are intuitive in non-interacting
theory but ``unnatural" due to gauge symmetry constraints.
These attempts have occasionally met remarkable successes. For example, although
the gluon spin is not a gauge-invariant concept,
the gluons in the IMF have momentum predominantly along the 3-direction and
appear to be on-shell. A free gluon is defined to have helicity of either $+1$ or $-1$.
It is then a simple matter to define the gluon helicity density $\Delta g(x)$ as the density
difference of two different helicities. The total gluon polarization
$\Delta G = \int \Delta g(x)$ is a measurable quantity through hard scattering,
and therefore must be gauge invariant. On the other hand,
in the rest frame of the nucleon, the transverse component of the
gluon momentum is by no means negligible, and the projection of the gluon spin
along the 3-direction is no longer gauge invariant. Moreover, as a part of
the bound state, the gluons are definitely off-shell. An off-shell gluon, as in the gluon
propagator, is hard to define gauge-invariantly. Therefore, $\Delta g(x)$ can no longer be
interpreted simply as the bound state property of the nucleon, although it is a
matrix element of a gauge-invariant operator, nonlocal
and complicated in general framework~\cite{manohar}. Nevertheless, in the
light-cone gauge, it reduces to the simple gluon spin operator along
the 3-direction, reminiscent of the IMF physics.

The example of the IMF gluon helicity has motivated myriad attempts to define
parton momentum and orbital angular momentum (OAM). In particular,
there has been a recent attempt to completely reinvent the concept of gauge invariance
by proposing to decompose the gauge potential into the so-called
physical and pure-gauge parts, and defining various ``physical
observables" with the physical part~\cite{Chen:2008ag,Chen:2009mr}.
The standard textbook result that gluons carry about one-half of
the nucleon momentum has been challenged by the new definition
of the parton canonical momentum. There have been a number of
follow-up studies in the literature~\cite{Wakamatsu:2010cb,Wakamatsu:2011mb,Hatta:2011zs,Leader:2011za,Zhou:2011zzf}.

In a different context, there have been many studies about the
parton transverse-momentum-dependent (TMD) factorization ``theorems''
in the literature. In some cases,
these theorems can be argued or even proved and the relevant distributions
are gauge invariant~\cite{tmdfac}. In many other cases, however, such factorizations
are simply model assumptions, and do not meet the requirement of
gauge symmetry. In particular, when the parton transverse momentum is fully
taken into account, the twist-expansion is often lost in the sense that
all twists are now involved. There seems to be no systematic
expansion since there are no arguments as to why
extra gluons with transverse polarizations should not be included in the
calculations~\cite{nontmdfac}.

In this paper, we analyze the constraints of gauge symmetry
on bound state properties and parton physics. In Sec. II, we make some general remarks about gauge symmetry,
particularly on how to maintain gauge symmetry in a fixed-gauge calculation.
We emphasize that the wave function is a gauge-dependent concept and that a gauge
symmetric-matrix element emerges only when the operator is
manifestly gauge symmetric, modulo the so-called BRST exact operators and the
equations-of-motion operators in the covariant gauge. In Sec. III, we analyze the gauge symmetry in parton physics, particularly in the case of gluon polarization. We point
out that while gluons appear as on-shell radiation in the IMF, they are off-shell and have
transverse momentum as a part of the bound state. We introduce the concept of gauge-invariant extension
(GIE) to establish a connection between parton physics and gauge symmetry.
In Sec. IV, we discuss the GIE of the Coulomb gauge in connection with the work
of Chen et al. We comment that the result of Wakamatsu's calculation of the scale evolution
for the gluon spin is in fact a light-cone gauge result~\cite{Wakamatsu:2011mb}. We will also discuss
the spin and OAM in the case of photon radiation
fields in optics. We conclude the paper in Sec. V.

\section{General Remarks about gauge symmetry}

In this section, we make some general remarks about gauge symmetry. We emphasize that
wave functions in non-relativistic quantum mechanics are gauge dependent, and the mechanical momentum of charged particles corresponds to the covariant derivative in quantum theory. We argue that in atomic systems, because
of the non-relativistic approximation, the partial derivative is actually gauge invariant to
leading order in $v/c$ expansion. We then consider the gauge symmetry in quantum field theories, reminding the reader
why gauge choice is a necessity and how a gauge-invariant result emerges from a fixed-gauge
calculation. We also comment on the relationship between gauge symmetry and Lorentz symmetry. These remarks are
by no means novel, and many are present in textbooks or papers. However, they lay out an important
general framework for gauge symmetry and will be useful for the discussions in the remainder of the paper.

\subsection{Gauge symmetry in non-relativistic quantum mechanics}

Let's first consider gauge symmetry in non-relativistic quantum theory with electromagnetic
interactions. For a charged particle moving in an external electromagnetic field,
the Hamiltonian is,
\begin{equation}
    H = \frac{(\vec{P}-e\vec{A})^2}{2m} + e\phi \ ,
\end{equation}
where $\vec{P}$ is the canonical momentum and we have customarily called $A^0 \equiv \phi$. It has the following
eigenvalue problem,
\begin{equation}
    H\psi_n(\vec{r}) = E_n \psi_n(\vec{r})\ ,
\end{equation}
with energy eigenvalues $E_n$ and energy eigen functions $\psi_n(\vec{r})$.

Under a time-independent gauge transformation,
\begin{equation}
    A^\mu(\vec{r}) \rightarrow {A^\mu} '(\vec{r})= A^\mu(\vec{r}) + \partial^\mu \chi(\vec{r})  \ ,
\end{equation}
we obtain a new Hamiltonian,
\begin{equation}
   H'= \frac{(\vec{P}-e\vec{A}')^2}{2m} + e\phi\ ,
\end{equation}
which is manifestly different. However, it has the same
quantum mechanical eigen energy,
\begin{equation}
  H' \psi_n'(\vec{r}) = E_n \psi_n'(\vec{r}) \ ,
\end{equation}
where $\psi_n' = e^{ie\chi(\vec{r})} \psi_n(\vec{r})$ is the eigen function in the
new gauge.
Thus, while the charged particle probability density,
\begin{equation}
  \rho_n(\vec{r})  = \psi_n^*(\vec{r})\psi_n(\vec{r}) \ ,
\end{equation}
is gauge invariant, the expectation value of the charged particle canonical momentum $\vec{P}$ (which is quantized as a
partial derivative)
is not:
\begin{equation}
   \left \langle \psi_m'\left|\vec{P}\right|\psi_n'\right\rangle
=    \left\langle \psi_m\left|\vec{P}\right|\psi_n\right\rangle  +e\left\langle \psi_m\left|\nabla \chi(\vec{r})\right|\psi_n\right\rangle \ .
\end{equation}
Thus, the proof that the canonical momentum matrix element is gauge invariant in Ref. \cite{Leader:2011za}
is likely incorrect. Actually, in Ref. \cite{Leader:2011za} the gauge transformation leaves the physical states invariant, but this is not true according to Eq. (5): wave function is a not a gauge-independent quantity.
It is the matrix element of the {\it mechanical momentum} $\vec{p} = \vec{P}-e\vec{A}$ that is gauge
invariant:
\begin{equation}
 \left\langle \psi_m'\left|\vec{P} - e\vec{A}'\right|\psi_n'\right\rangle
=    \left\langle \psi_m\left|\vec{P}-e\vec{A}\right|\psi_n\right\rangle \ ,
\end{equation}
which corresponds to the covariant derivative in quantization, $\vec{p} = \vec{P}-e\vec{A}\equiv -i\vec{D}$.

In Feynman's lectures on physics, he gave a simple example to demonstrate that it is
the covariant derivative that corresponds to the mechanical momentum in quantum
mechanics~\cite{feynmanlectures}: Consider a charged particle near a solenoid which
has a quantum mechanical wave function $\psi(\vec{r},t)$. The solenoid does not have
any current in the beginning. At some point in time, a current passes through the solenoid and
a stable magnetic field is established. In the process, the particle gets a momentum
kick because a changing magnetic field induces an electric field which exerts a force
on the charged particle. From the Schr{\"o}dinger equation, we know that 
the wave function must be continuous in time. Therefore, the momentum kick on the particle cannot be obtained from the partial derivative acting on the wave function, which must be
also continuous. Instead, this momentum kick comes from the establishment of the vector
potential $\vec{A}$ in the system.

\subsection{Atom systems and the Coulomb gauge}

Atoms are, to a very good approximation, non-relativistic systems.
In such systems, the charge effects are much larger than current
interactions, and the dominant interaction is the static Coulomb force.
Let us establish a power counting in terms of the velocity
$v$ and electromagnetic coupling $e$~\cite{Bodwin:1994jh}. The fine structure
constant $e^2/4\pi = \alpha_{\rm em}$ is a small quantity, and so
is the typical velocity of the charged particle $v/c \sim \alpha_{\rm em}$.
We will show through power counting that the canonical momentum and
canonical OAM are approximately gauge invariant.

In non-relativistic systems,
\begin{equation}
    B/E \sim v/c \ .
\end{equation}
We are going to use this to establish a power counting for the four-vector potentials.
According to the definition:
\begin{equation}
   \vec{E} = - \nabla \phi - \frac{\partial \vec{A}}{\partial t}, ~~~ \vec{B} = \nabla \times \vec{A} \ ,
\end{equation}
since the electrostatic interaction $e\phi$ is order of $\alpha^2_{\rm em}$ (recall the Bohr energy),
we have
\begin{equation}
  \phi \sim e\alpha_{\rm em}; ~~~~ E\sim {ev\alpha_{\rm em}} \ .
\end{equation}
In principle, we can choose $\vec{A}$ to be of the same order of magnitude as $\phi$. However, since $\vec{B}$ is suppressed
by a factor of $v/c$ relative to $\vec{E}$, the transverse part of $\vec{A}$
must be suppressed by the same factor. To have a natural power counting, one should work only in the
class of gauges in which the longitudinal part of $\vec{A}$ is similarly suppressed. One gauge
that satisfies this condition is the Coulomb gauge,
\begin{equation}
          \nabla \cdot \vec{A} = 0 \ .
\end{equation}
which eliminates the longitudinal part of the gauge potential entirely. Thus the
Coulomb gauge is a natural choice for non-relativistic systems.

From $B/E \sim v/c$, the counting for the vector potential in Coulomb gauge is
\begin{equation}
   \vec{A} \sim  ev\alpha_{\rm em} \ .
\end{equation}
Thus $e\vec{A}$ in the Coulomb gauge is $\alpha_{\rm em}^2$ suppressed relative to the canonical
momentum $\vec{P} = -i\nabla \sim v$. Therefore,
\begin{equation}
    \vec{p} = \vec{P} + {\cal O}(v\alpha^2_{\rm em}) \ ,
\end{equation}
and the canonical momentum is gauge invariant to the relative order $\alpha^2_{\rm em}$. This explains
why the canonical momentum and canonical OAM are useful concepts in atomic
systems.

Of course, one can choose to work in a gauge in which $e\vec{A}$ is not suppressed at all
relative to $\vec{P}$. Then the atomic wave functions will be different, and the physical
momentum and OAM must follow from the mechanical momentum and mechanical
OAM.

In QCD bound states, such as the nucleon, the above power counting fails.
In particular, the color magnetic field $\vec{B}^a$ is not suppressed relative to the
color electric field $\vec{E}^a$. Moreover, the quark motion is entirely relativistic. Therefore,
the $e\vec{A}$ and $\vec{P}$ have the same order of magnitude. The Coulomb
gauge is no longer special to keep the power counting natural. In factorization
theorems involving parton transverse momentum, it is often assumed that parton
canonical momentum is much larger than the transverse gauge potential in the light-cone gauge.
There is no known physical basis for this assumption.

\subsection{Quantization of QED and gauge symmetry}

Here we consider gauge-symmetry constraints in quantum field theories. For simplicity,
we take the U(1) gauge theory with a spin-1/2 fermion, such as quantum electrodynamics (QED), as an example.
The Lagrangian density is
\begin{equation}
   {\cal L} = \overline{\psi}(i\gamma_\mu D^\mu - m)\psi - \frac{1}{4}F^{\mu\nu}F_{\mu\nu} \ ,
\end{equation}
where $\psi$ is the electron's Dirac field and $A^\mu$ is the $U(1)$ gauge potential.

$A^\mu$ has two physical degrees of freedom; it also mediates static electric and magnetic
interactions among charged particles. However,
$A^\mu$ does have spurious degrees of freedom that must be constrained in order to
have a finite photon propagator. Thus, quantization of a gauge theory requires
gauge fixing, which ironically breaks the gauge symmetry. Although the symmetry
is no longer manifest after quantization, the physical observables in terms of
the physical degrees of freedom shall be independent of the spurious gauge dynamics.
This can be verified either through residual gauge symmetry after gauge fixing
or through comparing calculations in different gauges.

Different gauge fixings often lead to different Hilbert spaces, and the physical states
are gauge dependent, just as in the case of non-relativistic systems described
in the previous subsection.
For instance, in the Coulomb gauge, all quanta in the Hilbert space
are physical degrees of freedoms, whereas in the Weyl gauge $A^0=0$, the longitudinal polarization
of the photon is allowed, and the physical states are chosen through
Coulomb's law constraint $(\nabla \cdot \vec{E} - \rho)|\Psi_{\rm phys}\rangle=0$. In the
covariant Lorentz gauge, all four polarizations of the photons are allowed and the physical
Hilbert space is constrained by Lorentz gauge condition $\partial_\mu A^\mu|\psi_{\rm phys}\rangle=0$.
However, due to the residual gauge freedoms, the physical state is still not unique.
It can contain non-physical quanta.

Despite the fact that all Green's functions and wave functions are gauge dependent,
the poles of the Green's functions and scattering S-matrix are gauge invariant.
Usually, the proof is done using residual gauge symmetry. Here we are concerned with the
physical matrix elements in asymptotic states
$\langle \psi |O|\phi\rangle$~\cite{Leader:2011za}. In order for them to be gauge invariant, we have to study
general gauge symmetry transformation. Let's assume a superoperator $G$ (in the sense that it is capable of transforming states from one Hilbert space to other states in a different one) which brings the physical
states from one gauge to another,
\begin{equation}
                    |\psi'\rangle = G|\psi\rangle \ .
\end{equation}
Then the matrix element is invariant under gauge transformation,
\begin{equation}
                  \langle \psi'|O(A')|\phi'\rangle =  \langle \psi|O(A)|\phi\rangle \ ,
\end{equation}
only if $O(A')= O(A)$ is gauge invariant; that is, it has the same dependence on the gauge
potential as in $O$.

The requirement for matrix elements to be gauge invariant is much simpler to see in the path-integral
formulation~\cite{Hoodbhoy:1999dr}. The sufficient condition again is that the operator has to be
manifestly gauge invariant.

Of course, it is not necessary to have gauge-invariant operators in order for
its physical matrix elements to be gauge invariant. In covariant quantization,
it is known that the equations-of-motion operators and the so-called BRST-exact
operator do vanish in the physical states~\cite{Collins:1984xc}. For an example
of the energy-momentum tensor, see Ref~\cite{Ji:1995sv}.
It is also taken as true that the gauge-noninvariant total derivative operators do vanish in well-localized
states. However, in plane wave states, the total derivative operators do contribute
to the non-forward matrix elements. Thus one has to be careful with the total derivative operators,
particularly when they are gauge dependent.

The matrix element of a gauge-dependent operator usually vanishes when averaged over all gauges,
just like the velocity vector vanishes when summed over contributions from all frames~\cite{kfliu2011}.

\subsection{Lorentz symmetry}

According to Einstein's theory of special relativity, all physical quantities should be
Lorentz tensors, i.e., when coordinates transform, they behave either as scalars,
four-vectors, or high-order tensors. Therefore, when making a frame transformation,
one easily obtains the physical quantities in one frame from another.
In gauge theories, because not all four components of $A^\mu$ are physical,
calculations are found not to be manifestly Lorentz covariant. For example,
the gauge choices often break Lorentz symmetry when the gauge conditions are
frame dependent. However, for local
gauge-invariant quantities, Lorentz symmetry is manifestly kept in calculations~\cite{Ji:1997pf}.
For nonlocal operators, one has to introduce gauge links, which often
involve geometrical lines. The transformation of such a line is usually complicated, and
Lorentz symmetry transformation of nonlocal operators often becomes
intractable. In this case, one considers the subclass of Lorentz transformation
which leaves the line invariant.

In the case of noncovariant gauges, calculations in different frames often correspond
to different gauge choices. Therefore, gauge symmetry of the calculations is a
necessary condition to maintain Lorentz symmetry, or vice versa. We note that in quantum theory,
the wave function of a system is not a Lorentz invariant quantity---it changes with the coordinates.
Such a change is not just a kinematic transformation;
it involves a Lorentz boost which is usually interaction dependent. Thus boosting
a wave function is a highly nontrivial matter in a relativistic interacting theory.

\section{Extended Gauge Invariance in Parton Physics}

In high-energy scattering in QCD, partons are often used to
describe the scattering processes involving large momentum transfers~\cite{Feynman:1969ej}.
Total scattering amplitudes usually contain two factorized parts,
described as factorization theorems~\cite{Sterman:1994ce}. One part contains the product of
parton densities and/or fragmentation functions and/or soft functions.  As
far as the hard scattering is concerned, the hadrons
can be viewed as beams of non-interacting
partons, and the parton densities simply describe the properties of the beams.
The second part is the parton scattering cross section, which involves scattering
of on-shell colored particles. Feynman's parton language greatly
simplifies the description of high-energy scattering. The proof of the factorization
theorems is to show that the cross section has a systematic expansion
in terms of the inverse powers of large momenta and that the leading
factorized parts are naturally gauge invariant.

\subsection{Gauge symmetry of parton density and physics in the light-cone gauge}

\begin{figure}[hbt]
\begin{center}
\includegraphics[width=14cm]{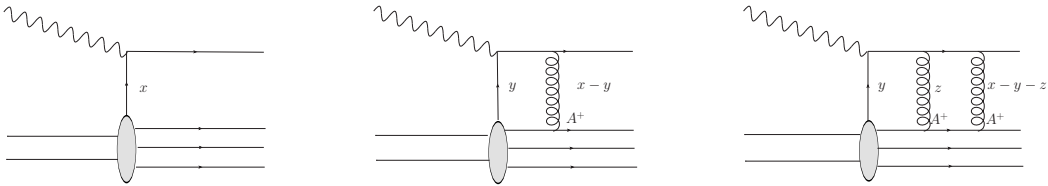}
\caption{Deep-inelastic scattering process in which the gauge invariance involving the
longitudinal quark mechanical momentum $xP^+$ is achieved through insertions of gluons with
longitudinal polarization $A^+$.}
\label{DIS}
\end{center}
\end{figure}

One of the keys to the parton language description of high-energy scattering is
gauge symmetry and the light-cone gauge,  $A^+=0$. [Here we use the light-cone
coordinates $A^\pm = (A^0 \pm A^3)/\sqrt{2}$, $A_\perp = (A^1,A^2)$, and it is well known
that the light-cone coordinates are equivalent to the use of IMF.]  The parton density can be
described as the matrix element of gauge-invariant operators in hadron states, and
since the cross section is gauge invariant by itself, this means that the on-shell parton
scattering cross section is also gauge invariant and can be calculated independently
in any gauge. It can also be seen that the parton cross sections involve only on-shell
parton scattering. The gauge invariance, however, masks the physical
property of the parton density operator. For example, the quark
longitudinal-momentum density distribution,
\begin{equation}
  q(x) = \frac{1}{2P^+} \int \frac{d\lambda}{2\pi}
  e^{i\lambda x} \langle PS|\overline{\Psi}(\lambda n)\gamma^+ \Psi(0)|PS\rangle \ ,
\end{equation}
where $n$ is a light-like four-vector $n^2=0$ and the nucleon momentum
is $P^\mu = (P^0, 0,0, P^3)$, with $P\cdot n=1$.
$\Psi(\xi)$ is a gauge-invariant quark field defined through
multiplication of a light-cone gauge link,
\begin{equation}
\Psi(\xi) = \exp\left(-ig\int^\infty_0 n\cdot A(\xi + \lambda n)d\lambda\right) \psi(\xi) \ .
\end{equation}
This gauge link ensures that whenever a partial derivative or canonical momentum of
colored quarks appear, the gauge potential $A^\mu$ must be present simultaneously
to make it a covariant derivative (mechanical momentum), $D^\mu = \partial^\mu + i gA^\mu$.
Indeed, taking its moments,
\begin{equation}
    \int x^{n-1} q(x) dx ~~~\sim~~~ n_{\mu_1} ...n_{\mu_n}
    \langle P|\overline{\psi}(0) \gamma^{\mu_1}iD^{\mu_2}... iD^{\mu_n} \psi(0)|P\rangle \ ,
\label{moments}
\end{equation}
we see that the parton momentum distribution refers to the gauge-invariant mechanical momentum!
The mechanical momentum structure is clearly seen through Feynman diagrams
in Fig.~\ref{DIS}: Gauge symmetry requires that a parton with
mechanical momentum $k^+ = xP^+$ includes the sum of all diagrams with
towers of longitudinal gluon $A^+$ insertions.

Simple parton physics emerges in the light-cone gauge $A^+=0$, where the light-cone
gauge link disappears and all the covariant derivatives in Eq. (\ref{moments})
become partial ones, i.e., the canonical momentum and the simple quark field become
physical. One can use the light-cone quantization to write:
\begin{equation}
  \psi_+(\xi) = \int \frac{dk^+ d^2k_\perp}{2k^+(2\pi)^3}
     \left[ d^\dagger(k^+,k_\perp) v(k^+,k_\perp) e^{i(k^+\xi^--\vec{k}_\perp\cdot \vec{\xi}_\perp)}
           + b(k^+,k_\perp) u(k^+,k_\perp) e^{-i(k^+\xi^--\vec{k}_\perp\cdot \vec{\xi}_\perp)}\right] .
\end{equation}
Inserting the above expression into the parton density, one finds for $x>0$,
\begin{equation}
   q(x) = \int \frac{d^2k_\perp}{2k^+(2\pi)^3} \langle P|b^\dagger(xP^+, k_\perp) b(xP^+,k_\perp) | P\rangle \ ,
\end{equation}
which is a genuine number-density distribution.

While the parton language simplifies the description of a scattering process
considerably, one does not have to choose the light-cone gauge $A^+=0$.
There always exists a natural gauge-invariant formalism
that embodies the factorization, such as the presence of the matrix elements of the
twist-two gauge-invariant operators.

In recent years, there have been many discussions on the so-called $k_T$-dependent factorizations.
The motivation is to take into account the effects of parton transverse momentum in the scattering
processes. In some cases where there is a systematic power counting for the relevant
physical observable, the factorization is gauge invariant~\cite{tmdfac}. However, in many other cases,
the factorization is basically a model assumption which cannot be justified because
other contributions of the same order have been neglected and the result is not gauge
invariant~\cite{nontmdfac}.

\subsection{Gluon spin}

It is well known that in gauge theories, the spin operator of a gauge particle, defined as
\begin{equation}
  \vec{S}_g = \int d^3\xi \vec{E}\times \vec{A} \ ,
\end{equation}
is not a gauge-invariant quantity~\cite{landau}, because of the manifest
dependence on the gauge potential $\vec{A}$. This applies for photons in QED and gluons in QCD.
However, its helicity, i.e., the projection of $\vec{S}_g$ along the direction of momentum, is
gauge invariant. This can be easily seen from the vanishing of the gauge-dependent term $(\vec{\nabla} \times \vec{\nabla} \chi) \cdot \vec{E}$. Helicity is also a projection of the total angular momentum along the direction
of motion, as the orbital part obviously does not contribute. The helicity of a vector
gauge particle is defined to be $+1$ or $-1$, corresponding to right-handed or
left-handed polarization, respectively.

While helicity is a useful gauge-invariant concept, it should not be confused with
spin projection. For example, the spin vector $\vec{S}_g$ calculated along the $\perp$-direction for a photon
moving in the $3$-direction definitely depends on the choice of the wave function
or gauge. Many concepts developed in radiation physics use helicity as a starting point.
For example, the well-known electromagnetic multipoles are constructed with Wigner rotation matrices
applying to plane wave states with good helicities~\cite{landau}.

A nucleon with large $3$-momentum $P_3\rightarrow \infty$ is made of
off-shell (virtual) constituents with large $3$-momenta, $xP_3$, as well. In high-energy scattering,
only this $3$-momentum is important and the transverse component and off-shellness can be neglected
because they produce subleading effects in the parton
scattering cross section. Therefore, as far as the hard scattering process is concerned,
the nucleon appears to be made of a beam of massless {\it on-shell} partons (quarks and gluons)
moving in the same direction (called equivalent-photon or Weizs\"acker-Williams approximation in QED~\cite{jackson}).
For such a beam of gluons, the helicity is a good quantum number. The total beam helicity---the difference between the numbers of particles with positive and negative helicity---is
also a physical observable. Thus, we have the gauge-invariant gluon helicity distribution $\Delta g(x)$,
\begin{equation}
  \Delta g(x) = g_+(x) - g_-(x) \ ,
\end{equation}
where $g_\pm(x)$ is the density of gluons with momentum $x$ and helicity $\hat{P}\cdot \vec{S}_g = \pm 1$.

However, in the bound state nucleon, the gluons are virtual quanta which not only involve
transverse polarization but also longitudinal polarization. They are
off-shell particles whose wave function cannot be defined in a gauge-invariant way. Thus, the
gluon helicity distribution formally is a matrix element of a
complicated nonlocal operator~\cite{manohar},
\begin{equation}
  \Delta g(x) = \frac{i}{2x(P^+)^2} \int
  \frac{d\lambda}{2\pi} e^{ix\lambda} \langle PS| F^{+\alpha}(\lambda n) L_{[\lambda n^-,0]}\tilde F^+_{~\alpha} (0)|PS\rangle \ .
\end{equation}
where $\tilde F^{\alpha\beta} = 1/2 \epsilon^{\alpha\beta\mu\nu}F_{\mu\nu}$ and $L_{[\lambda n^-,0]}$ is a light-cone gauge link.
The need for gauge invariance dictates that it must be the gluon field strength $F^{\mu\nu}$ that appears,
and the gauge link ensures cancellation of the gauge transformation factors $U$ from
two different spacetime points. In the light-cone gauge, $\Delta g(x)$ becomes the matrix element of a local operator. The QCD expression only ensures that in the IMF and light-cone gauge,
it reduces to the simple gluon helicity counting,
\begin{equation}
  \Delta g(x) = \int \frac{d^2k_\perp}{2k^+(2\pi)^2} \langle P|a^\dagger_{+1}(xP^+, k_\perp) a_{+1}(xP^+,k_\perp) - a^\dagger_{-1}(xP^+, k_\perp) a_{-1}(xP^+,k_\perp) | P\rangle  \ .
\end{equation}
In light-cone quantization with $A^+=0$, an on-shell gluon has two physical polarizations, described
by the transverse components of the gauge potential $A_\perp$.
The matrix element is, however, independent of the momentum
in the 3-direction and can be taken in the rest frame of the nucleon.  However, in such a frame, the gluon parton's
transverse momentum definitely cannot be neglected, neither can its off-shellness. For these gluons, their spin
projection in the 3-direction is no longer gauge invariant; one cannot define the
gauge-invariant gluon helicity in a simple fashion. Therefore, $\Delta g(x)$ does not have a straightforward physical interpretation as the property of the bound state.

Indeed, when integrating over all $x$, one gets the total parton helicity operator in a
longitudinally polarized nucleon,
\begin{equation}
    S_g^{\rm inv} = \frac{i}{2}\int \frac{dx}{xP^+} \int d^3\xi e^{ix\xi^-P^+} F^{+\alpha}(\xi^-)L_{[\xi^-,0]} \tilde F_{~~\alpha}^+(0) \ ,
\label{gspingie}
\end{equation}
which is a nonlocal operator, and again has no clear gauge-invariant physical interpretation.
$S_g$ has a simple interpretation only in the light-cone gauge: it reduces to the gluon spin,
\begin{equation}
          S_g^{\rm inv}\left|_{A^+=0}\right.= \int d^3\xi (\vec{E}\times \vec{A})^3 \ .
\end{equation}
Of course, as we have discussed in Sec. II, the same operator defined in other gauges
has entirely different bound-state matrix elements.

The above discussion makes clear that gluon spin is
not a gauge-invariant quantity, but in the light-cone gauge and IMF, the gluons
in a bound state may be regarded as on-shell and the gluon helicity becomes a useful concept and can
be measured experimentally (similar to electromagnetism~\cite{jackson}). Leaving this specific gauge and frame, this quantity involves longitudinally polarized gluons
and no longer has a clear physical interpretation, though it can be defined and calculated gauge-invariantly.
This motivates going beyond the traditional sense of gauge symmetry by introducing the
{\it gauge-invariant extension} (GIE) of some apparently gauge-dependent
quantities. A GIE of a gauge-noninvariant quantity is to generalize the fixed gauge result
to any other gauge. There are an infinite number of GIE
for a gauge-dependent quantity depending on the gauges in which one decides to
generalize. The gluon spin operator
in the light-cone gauge has its GIE in Eq. (\ref{gspingie})~\cite{manohar}.

\subsection{GIE of parton observables}

Parton physics emerges in the set-up of the IMF and light-cone gauge.
In a fixed gauge, not all results are naturally gauge invariant, as
gauge fixing often masks the gauge symmetry properties of both
gauge-invariant as well as gauge-noninvariant quantities.
This is similar to calculations in a fixed system of coordinates
where physical quantities often do not have apparent properties
of the underlying tensors.

One can make GIE for any partonic operators using the
same approach as in the case of the gluon spin, so that
in any other gauges, GIE partonic operators produce the same
matrix elements as in the light-cone gauge.

For example,
a GIE of the partial derivative in the light-cone gauge is
\begin{equation}
   i\partial^\perp_\xi = iD^\perp_\xi + \int^{\xi^-} d\eta^-  L_{[\xi^-,\eta^-]} gF^{+\perp}(\eta^-,\xi_\perp) L_{[\eta^-,\xi^-]} \  .
\end{equation}
In this way, the partial derivative in the
light-cone gauge becomes gauge invariant, and therefore one can talk
about ``gauge-invariant" canonical momentum and orbital angular
momentum. In fact, the canonical version of the
spin sum rule discussed in Refs.~\cite{Jaffe:1989jz,Bashinsky:1998if} and its scale
evolution~\cite{Hoodbhoy:1999dr} can be understood as ``gauge-invariant"
in this sense~\cite{jixiongyuan,Hatta:2011ku}.

In the same sense, one may regard all $k_T$-dependent factorization
result as gauge invariant.

\subsection{Problems with gauge-invariant extension}

Clearly, the GIE of an intrinsically gauge-noninvariant quantity is not naturally
gauge invariant. As such, its usage has a number of problems:

First, the operators resulting from GIE are in general nonlocal. Nonlocal operators do not usually have a
clear physical meaning in general gauges, although they usually do in a fixed gauge. It is also difficult
to perform calculations which are most practically done in the fixed gauge. In fact, all GIE of partonic
operators cannot be calculated in lattice QCD because they intrinsically involve the
real time which becomes imaginary after the Wick rotation. This, for example, is the
case for the total gluon helicity $\Delta G = \int dx \Delta g(x)$.

Second, GIE operators do not transform simply under Lorentz transformations. While local gauge-invariant
operators often have simple classifications in terms of representations of the Lorentz group, nonlocal
operators involve geometrical lines as gauge links which do not have simple Lorentz transformation
properties. Many gauge conditions are not Lorentz covariant.

Third, there are (infinite) many nonlocal operators that have the same quantum numbers. For instance,
one can merely change the path of a gauge link to get a new operator. All of these operators
can mix under scale evolution. Therefore, scale-mixing calculations become intractable~\cite{Hoodbhoy:1998yb}.

Finally, GIE operators are in general unmeasurable. So far, the only example
is offered in high-energy scattering in which certain partonic GIE operators may be measured.
For GIE operators in the Coulomb gauge, to which we will return in the next section,
there is no known physical measurement for any of them.

For the transverse-momentum-dependent factorization, many of the so-called factorization
formulae are intrinsically gauge dependent. There are contributions of the same order that have been neglected, for instance, contributions from additional transversely polarized gluons.
Here gauge dependence is a signal that there are legitimate contributions that have been left out.
Thus one makes little gain by claiming the result is gauge invariant through GIE.

\section{gauge-invariant extension of the Coulomb gauge}

The Coulomb gauge is a natural gauge choice in atomic physics, which as we discussed in Sec. II
preserves the power counting in non-relativistic effective theory.
In this section, one considers the GIE of the Coulomb gauge, i.e., extending a
Coulomb gauge result gauge-invariantly to other gauges. For a naturally
gauge-invariant quantity, this of course does not generate anything new.
However, if one believes that the Coulomb gauge potential $A^\mu_c$
is special, any operator made from it can be made gauge invariant
through GIE. As we shall see, this formalism is just what Chen et al. have
proposed as a new ``gauge-invariant" formulation~\cite{Chen:2008ag,Chen:2009mr}.
The formulation was in fact based on the well-known observation
that in a fixed frame, any electric field $\vec{E}$
can be separated into a longitudinal part $\vec{E}_{||}$ and
a transverse part $\vec{E}_{\perp}$. The transverse part of the electric field
and the total magnetic field can be generated from a transverse gauge potential
$\vec{A}_{\perp}$, and the longitudinal part can be generated from the combinations
of  $\vec{A}_{||}$ and $\phi$, and hence the gauge freedom. The Coulomb gauge
corresponds to choosing $\vec{A}_{||}=0$. Although $\vec{A}_{\perp}$ is now gauge
invariant, it is frame dependent. Therefore, gauge dependence
in this formalism is traded into frame dependence.

In this section, we will first define the GIE of the Coulomb gauge. Then
we argue that Chen et al.'s proposal is just the same. We show that the so-called
gauge-invariant anomalous dimension of the gluon spin operator
calculated in Ref.~\cite{Wakamatsu:2011mb} is, in fact, corresponding to the GIE of the gluon spin in
the light-cone gauge. Finally, in the last subsection, we discuss why the spin and orbital angular
momentum separation seems possible in optics.

\subsection{Coulomb gauge extension}

Let us consider how to make the GIE of the Coulomb gauge. For simplicity, we consider QED only and use a subscript
$c$ to denote quantities in this gauge. The gauge condition is
\begin{equation}
                    \vec{\nabla}\cdot \vec{A}_c = 0 \ ,
\end{equation}
and $A^0_c$ becomes a nondynamical variable
satisfying the constraint:
\begin{equation}
                  \nabla^2 A^0_c = -\vec{\nabla} \cdot \vec{E} = -\rho \ .
\end{equation}
Let us consider any Coulomb gauge operator $O_c$ as a function of $A_c^\mu$.
Its matrix element evaluated in the Coulomb gauge wave functions is
$\langle \psi_c|O_c|\phi_c\rangle$. Any such operator has a GIE
${\bf O} \equiv O_c(A^\mu_c)$ by replacing every $A_c^\mu$ in it by
\begin{equation}
             A_c^\mu = \frac{1}{\nabla^2}\partial^i F_{i\mu} \ .
\end{equation}
which depends on the gauge-invariant field $F_{\mu\nu}$.
${\bf O}$ is now manifestly gauge invariant as $A_\mu$ only appears in
the field strength tensor.
If we now specialize ${\bf O}$ in any gauge by letting $A^\mu$
satisfy a new gauge condition $g$, calling it ${\bf O}_g$,
its matrix element is then the same as that in the Coulomb gauge:
\begin{equation}
           \langle \psi_g|{\bf O}_g|\phi_g \rangle \equiv \langle \psi_c|O_c|\phi_c\rangle \ .
\end{equation}
Note that the states $|\psi_g\rangle$ and $|\phi_g\rangle$ are correspondingly obtained
in the same gauge $g$. In this sense, any operator in the Coulomb gauge $O_c$ is
``gauge invariant" through ${\bf O}$, although it is generally nonlocal in terms of
the non-gauge-fixed potential $A_\mu$.

\subsection{Chen et al.'s proposal}

Chen et al. claimed to have found a way to describe the spin of a gauge particle
through a new formulation of gauge symmetry~\cite{Chen:2008ag}. The first step in their theory is to
decompose the gauge potential $A^\mu$ into two parts:
\begin{equation}
  A^\mu = A^\mu_{\rm pure} + A^\mu_{\rm phys}  \ .
\end{equation}
The part $A_{\rm pure}^\mu$ carries all the gauge degrees of freedom and hence has
no physical consequence. The other part is ``physical" because it is
invariant under gauge transformations. In their choice, they
demand $\vec{A}_{\rm phys}$ to satisfy the divergence-free condition:
\begin{equation}
   \vec{\nabla}\cdot \vec{A}_{\rm phys} = 0 \ ,
\end{equation}
whereas $\vec{A}_{\rm pure}$ satisfies the curl-free condition:
\begin{equation}
  \vec{\nabla}\times \vec{A}_{\rm pure} = 0 \ .
\end{equation}
Under gauge transformation, one has
\begin{equation}
   \vec{A}_{\rm pure} \rightarrow  \vec{A}_{\rm pure}' =  \vec{A}_{\rm pure} + \nabla \chi; ~~~
   \vec{A}_{\rm phys} \rightarrow \vec{A}_{\rm phys}'  = \vec{A}_{\rm phys}  \ .
\end{equation}
Therefore, one can construct gauge-invariant observables
using $\vec{A}_{\rm phys}$, including for example, the gluon spin density, $\vec{E}\times \vec{A}_{\rm phys}$.

It is easy to see that this proposal is the same as the GIE of the Coulomb gauge.
If one is to make a calculation, the simplest way
is to choose,
\begin{equation}
\vec{A}_{\rm pure} = \vec{\nabla}\frac{1}{\nabla^2} \vec{\nabla} \cdot \vec{A} =  0 \ ,
\end{equation}
which is the same as the Coulomb gauge condition, $\vec{\nabla} \cdot \vec{A}=0$.
Then, $\vec{A}_{\rm phys}$ is the physical part of $A^\mu$ in the Coulomb gauge which also
satisfies the condition $\vec{\nabla} \cdot \vec{A}_{\rm phys}=0$. The QED lagrangian
contains only $\vec{A}_{\rm phys}$, and all calculations proceed in exactly the
same way as in the Coulomb gauge, and all results are naturally the same as those in
that gauge. In particular, the matrix element $\vec{E}\times \vec{A}_{\rm phys}$
is the same as  $\vec{E}\times \vec{A}$ calculated in the Coulomb gauge.

The procedure that Chen et al. proposed is to ensure calculations in any gauge involving
$\vec{A}_{\rm phys} = \vec{A} -  \vec{\nabla} \frac{1}{\vec{\nabla}^2}\vec{\nabla}\cdot \vec{A}$
yield the same result as that in the Coulomb gauge. This is the significance of ``gauge invariance"
associated with the matrix element of operators such as $\vec{E}\times \vec{A}_{\rm phys}$. In fact,
the propagator $\langle 0|T A^{\mu}_{\rm phys} A^{\nu}_{\rm phys}|0\rangle$, considered as gauge invariant and physical, computed in any other
gauge will yield the Coulomb gauge result:
\begin{equation}
\langle 0|T A^{\mu}_{\rm phys} A^{\nu}_{\rm phys}|0\rangle
= \frac{i}{k^2+i\epsilon} \left(-g^{\mu\nu} - \frac{k^\mu k^\nu}{\left((k\cdot \eta)^2-k^2\right)} + \frac{(k\cdot \eta)(k^\mu \eta^\nu + k^\nu
\eta^\mu)}{\left((k\cdot \eta)^2-k^2\right)}\right) \ ,
\end{equation}
where $\eta^\mu$ is a temporal vector $\eta=(1,0,0,0)$. This confirms the conclusion that Chen et al.'s
approach is the same as the GIE of the Coulomb gauge.

Of course, as discussed above, working in a fixed frame allows one to isolate the transverse and longitudinal
parts of the electric field and gauge potential, and hence allows one to decompose total momentum into
the canonical-like contributions~\cite{Chen:2009mr}. All calculations involving $A_{\perp}$ are
gauge invariant. However, while the gauge part of the momentum appears simpler, the
charged particle's momentum does not have a simple physical interpretation as discussed Sec.II.
Furthermore, $A_{\perp}$ does not transform simply under Lorentz transformation,
and quantities that are not naturally gauge symmetric will have complicated momentum dependence coming from the
dynamical origin.

\subsection{Wakamatsu's approach }


Following the work by Chen et al., Wakamatsu~\cite{Wakamatsu:2010cb} proposed to generalize the procedure of separating the gauge and gauge-invariant parts of the potential such that one can impose alternative
conditions on the latter and still maintain the gauge symmetry of Chen et al.'s decomposition. He claimed that in this way one can connect different decompositions as well as different frames. Unfortunately, many of the
claims in the paper are incorrect. His procedure is not gauge invariant, and different conditions
on the ``gauge-invariant" part produce different results. Moreover, when the separation is not Lorentz
covariant, the discussion on frame independence is also incorrect. Here we focus solely on the
claim that the evolution of the gluon spin in this decomposition is gauge invariant~\cite{Wakamatsu:2011mb}.
His definition of the gluon spin is not that of Chen et al., but the GIE of the light-cone gauge.
Therefore, his anomalous dimension result must coincide with the light-cone gauge calculation of
the gluon spin operator, and this is exactly what he found. So his calculation does not at all support the claim
that the gluon spin he defined is gauge invariant.

The first step in Wakamatsu's work is also to decompose the gauge field $A^\mu$ into $A^\mu_{\rm pure}$ and $A^\mu_{\rm phys}$. In the paper~\cite{Wakamatsu:2011mb}, $A^\mu_{\rm phys}$ is defined by imposing the light-cone condition $n\cdot A_{phys}=0$, where $n^2=0$. The propagator $\la0|T[A^\mu_{\rm phys}A^\nu_{\rm phys}]|0\ra$ computed in any gauge just yields the light-cone result:
\begin{eqnarray}
\la0|T[A^{a,\ \mu}_{\rm phys}(x)A^{b,\ \nu}_{\rm phys}(y)]|0\ra&=&\int{d^4k\over(2\pi)^4}e^{ik\cdot(x-y)}\left({-i\delta^{ab}\over k^2+i\epsilon}\right)\left(g^{\mu\nu}-{n^\mu k^\nu+n^\nu k^\mu\over n\cdot k}\right)  \ .
\end{eqnarray}
The gluon spin operator is constructed from $A^\mu_{\rm phys}$:
\begin{eqnarray}
M^{+12}_{\rm g-spin}&=&2\mbox{Tr}[F^{+1}A^2_{\rm phys}-F^{+2}A^1_{\rm phys}] \ .
\label{gluonspin1}
\end{eqnarray}
This, of course, is the same spin operator as in the light-cone gauge, which can be made gauge
invariant by the procedure discussed in the previous section. Once this operator was generalized to the other gauges,
Wakamatsu calculated the one-loop anomalous dimension in the Feynman gauge, which turned out to be the same as that
obtained within the light-cone gauge by Hoodbhoy et al.~\cite{Hoodbhoy:1999dr}. Nevertheless, this result is hardly novel from the GIE point of view.

The GIE of the gluon spin operator from the light-cone gauge is realized by choosing
\begin{eqnarray}
A^\mu_{\rm phys}(\xi) &=& \int^{\xi^-}d\eta^-\ L_{[\xi^-,\eta^-]}\ F^{+\mu}(\eta^-,\xi_\perp)\ L_{[\eta^-,\xi^-]} \ ,
\end{eqnarray}
where $L$'s are light-cone gauge links. In this sense, the gluon spin operator constructed in Eq.~(\ref{gluonspin1}) is gauge invariant, and its matrix elements in any other gauge should be just the same as they are in the light-cone gauge.
This is also true for its anomalous dimension.
We have verified our conclusion by explicit calculations
in several choices of gauges. At the lowest order of $A^\mu_{\rm phys}$, we calculated $\Delta\gamma^{(0)}_{Gq}$ at one-loop level in general covariant and noncovariant gauges. All the additional integrals, with respect to the Feynman gauge result, cancelled each other out exactly. Therefore, Wakamatsu's result in the Feynman gauge cannot be any thing other than that in the light-cone gauge.

Since Wakamatsu's calculation does not agree with Chen et al.'s result in the Coulomb gauge~\cite{Chen:2009mr},
he suspected that the latter was calculated incorrectly~\cite{Wakamatsu:2012ve}. Actually, we confirmed the Coulomb gauge calculation. The difference
is due to different GIEs. Wakamatsu's generalized $A_{\rm phys}$ does not lead to a gauge-invariant result, which is contrary to his claim.

\subsection{Photon spin and orbital angular momentum in atomic physics}

It has often been claimed that the photon spin and orbital angular momenta in QED can be separately defined and
measured, and therefore must be individually gauge invariant. This has served as another important
motivation to look for a gauge-invariant definition of the gluon spin. In this subsection, we discuss the examples in optics, pointing out that this is possible
only for optical modes with a fixed frequency.

We first recall a bit of history about photon's angular momentum. For a circularly polarized plane wave, R. Beth~\cite{Beth1936} was the first to measure its spin angular momentum by measuring the torque exerted on the quartz wave plate it passed through. As we explained above, this is simply the helicity and is gauge invariant in theory.
In 1992, L. Allen et al. pointed out that Laguerre-Gaussian laser modes also have a well-defined orbital angular momentum~\cite{Allen1992}. Based on this, several experiments~\cite{O'Neil2002, Anderson2006, Leach2002} have been set up to observe and measure the orbital angular momentum of a Laguerre-Gaussian photon.

For radiation field with $e^{-i\omega t}$ time-dependence, using the Maxwell equation:
\begin{eqnarray}
\vec{B}&=&-{i\over\omega}\nabla\times\vec{E} \ ,
\end{eqnarray}
one always has the gauge-invariant decomposition:
\begin{eqnarray}
\vec{J}&=&{1\over2}\int d^3\xi\ [\vec{\xi}\times(\vec{E}^*\times\vec{B}+\vec{E}\times\vec{B}^*)]\nonumber\\
&=&-{i\epsilon_0\over \omega}\int d^3\xi [\vec{E}^*_i(\vec{\xi}\times\nabla)\vec{E}_i+\vec{E}^*\times\vec{E}]  \ .
\end{eqnarray}
The two terms may be identified as the orbital and spin angular momentum, respectively, and are separately
gauge invariant. In particular, this is true for photon
electric and magnetic multipoles which are often used in
transitions between atomic or nuclear states. The photon orbital angular momentum can be defined
without referring to the gauge potential at all~\cite{jackson}.

It can be easily checked from the above equation that the spin equals $\mp1$ for left-handed or right-handed circularly polarized light, and 0 for linearly polarized light. In paraxial approximation, a Laguerre-Gaussian mode with azimuthal angular dependence of $\mbox{exp}(il\phi)$ is an eigen mode of the operator $L_z=-i\partial/\partial\phi$, and carries an orbital angular momentum of $l\hbar$. It is remarkable that the experimentalists are able to find ways to detect the effects of the orbital
angular momentum alone~\cite{O'Neil2002, Anderson2006, Leach2002}.


Clearly, the above procedure only applies for a specific type of radiation field. In the case
of QCD, the gluons in the nucleon cannot be of this type. In particular,
they are off-shell and do not satisfy the on-shell equations of motion. The best one can
do is to go to the IMF where the gluons appear as on-shell radiation; in this way, the gluon helicity
and orbital angular momentum could be defined and measured ``naturally" in the light-cone
gauge. Thus we are back to the discussions made in the previous section.

\section{Conclusion}

This paper examines the constraints of gauge invariance on
physical observables in a gauge theory. We advocate the
standard textbook view that the naturally gauge-invariant (local)
operators must be manifestly invariant under gauge transformations.
Thus, the spin of gauge particles, canonical momentum, and canonical OAM are not
strictly-speaking gauge invariant. Only in special contexts, it makes sense
to talk about this. These special cases arise 1) in non-relativistic
quantum systems such as atoms where the Coulomb gauge is a natural
choice; 2) in radiation fields where the time variation is in harmonic
oscillation (such as the electromagnetic multipoles and Laguerre-Gaussian laser modes); 3) in high-energy hadron scattering in which the gauge
partons appear to have two physical polarizations and
be on-shell. In the last case, the simple physics in the IMF
and light-cone gauge cannot be extended to that of the bound state
because the same quantity does not have a simple gauge-invariant
interpretation in the rest frame of the nucleon. In fact, there
is no gauge-invariant notion of the gluon spin contribution
to the spin of the nucleon. The IMF frame result can be translated
into the physics of bound state only in the light-cone gauge.

Of course, high-energy scattering has provided great tools to learn about
parton physics. Thus, it is possible to probe, for example, the canonical
momentum and OAM of the partons in the IMF and light-cone gauge~\cite{jixiongyuan,Hatta:2011ku}.
However, they cannot be taken as gauge-invariant properties of bound states,
contrary to many assertions made in the literature so far. Also, many of the TMD
factorization theorems are practically model assumptions, which must be
taken with caution~\cite{nontmdfac}.

Thus, the efforts of trying to generalize the gauge symmetry from a fixed
gauge calculation is of limited use. In particular, generalizing
the Coulomb gauge results as ``gauge-invariant" in the case
of QCD appears to be neither physically attractive nor experimentally relevant~\cite{Ji:2010zza,Ji:2009fu}.

This work was partially supported by the U.
S. Department of Energy via grants DE-FG02-93ER-40762
and a grant (No. 11DZ2260700) from the Office of
Science and Technology in Shanghai Municipal Government. One of the authors, Yang Xu, would also like to thank the China Scholarship Council for their award through the State Scholarship Fund to pursue her studies abroad, and to the University of Maryland for its hospitality and support. We thank E. Leader for comments on the initial version of the manuscript.

\end{document}